\documentclass[conference]{IEEEtran}
\IEEEoverridecommandlockouts
\usepackage{cite}
\usepackage{amsmath,amssymb,amsfonts}
\usepackage{algorithmic}
\usepackage{graphicx}
\usepackage{textcomp}
\usepackage{xcolor}
\usepackage{booktabs}

\begin{document}

\title{A Simple Data Exfiltration Game}

\author{\IEEEauthorblockN{Tristan Caulfield}
\IEEEauthorblockA{\textit{Department of Computer Science} \\
\textit{University College London}\\
London, United Kingdom \\
t.caulfield@ucl.ac.uk}
}



\maketitle

\begin{abstract}
Data exfiltration is a growing problem for business who face costs related to the loss of confidential data as well as potential extortion.  This work presents a simple game theoretic model of network data exfiltration.  In the model, the attacker chooses the exfiltration route and speed, and the defender selects monitoring thresholds to detect unusual activity.  The attacker is rewarded for exfiltrating data, and the defender tries to minimize the costs of data loss and of responding to alerts.
\end{abstract}


\section{Introduction}

Data exfiltration is becoming an increasingly worrying cybersecurity problem.  Ransomware attacks are starting to no longer just encrypt but also to exfiltrate data for what is called double extortion: demanding payment to decrypt the data, and further payment to refrain from releasing the stolen data publicly.  Losses to companies from data exfiltration can be significant, including lost of reputation or business, disruption, loss of confidential information, regulatory fines, and more.

Attackers attempt to move stealthily and remain undetected while defenders employ tools that monitor network traffic for unusual activity which may indicate a breach or exfiltration attempt.
These tools can alert response staff or automatically block traffic when an anomaly is detected; however, these actions come with a cost. Investigating alerts or disrupting service can be costly.  If there are too many false positives from the tool, it can impose large costs on an organization.

In this work we present a simple model that explores the trade-offs of both attacker and defender.  The attacker chooses the route that will be used to exfiltrate data as well as the speed, and the defender must select alert thresholds and try to balance the costs of data loss against the costs of monitoring.  We build a straightforward game theoretic model and then explore how its various parameters affect system performance.

\section{Related Work}

Game theory applied to network exfiltration has been studied before in a number of papers.
In \cite{mc_carthy_data_2016}, McCarthy et al develop an approach to detecting data exfiltration over DNS queries.  They use POMDPs to learn optimal policies for acting based on observations of information from the network.

Durkota et al \cite{durkota_optimal_2017} look at data exfiltration in networks with different types of attackers who have different knowledge of the system.  This work also uses POMDPs and the authors develop algorithms to find optimal defense strategies with Stackelberg equilibria.

Nguyen et al \cite{nguyen_stackelberg_2017} find optimal strategies for allocation resources to detecting botnet exfiltration attacks, using a zero-sum Stackelberg game.  In this model, exfiltration can happen along single or multiple paths.

In \cite{mitchell_game_2018}, Mitchell and Healy look at a slightly larger picture and build a game theoretic model of many stages of computer network exploitation.  This includes many phases, such reconnaissance, lateral movement, as well as exfiltration.

More recent related work looks at extortion ransomware.  While not about detection, Meurs et al \cite{meurs_deception_2024} build a signalling game to approach the problem of information asymmetry in ransomware negotiations where the victim does not know whether data has been exfiltrated or not.

\section{Setup}

For this work, we want to create a simplified model of a computer network and exfiltration detection system.  We want to capture network structure, a choice of routes and exfiltration speeds for the attacker, and a choice of detection thresholds for the defender.
We assume the attacker is already in the network, successfully hidden from the defender, and only detectable through their exfiltration attempt.

We use a directed graph to represent the possible routes for data exfiltration on the network.  We call $l \in links$ a link from $l_\mathrm{src}$ to $l_\mathrm{dest}$.  A route is a tuple of $m$ links such that $l_{1_\mathrm{dest}} = l_{2_\mathrm{src}}, \dots, l_{m-1_\mathrm{dest}} = l_{m_\mathrm{src}}$. 

\begin{figure}
    \centering
    \includegraphics[width=.7\columnwidth]{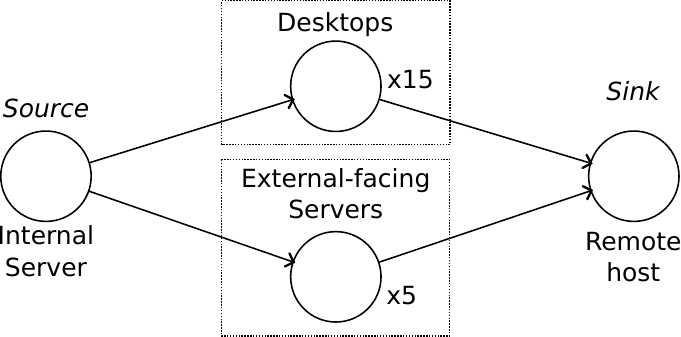}
    \caption{A simple network configuration. Attackers aim to exfiltrate data from source to sink.}
    \label{fig:network1}
\end{figure}

Figure~\ref{fig:network1} shows an example network structure.  Routes must start at a source, representing, for example, a server where confidential data is stored; they must finish at a sink, representing a host outside of the network that is the destination for exfiltrated data.  The total amount of data available for exfiltration is given by $D$.

Each link $l$ has an associated distribution $\mathrm{dist}_l$ with mean $\mu_l$ and standard deviation $\sigma_l$.  This distribution represents the normal or historical behaviour of the network traffic on this link.  Activity by an attacker may increase the traffic on a link. For each link, the defender selects a threshold, $T_l$, which represents the alert level in their exfiltration detection system.  If the activity on a link exceeds $T_l \sigma_l$, an alert is triggered, leading to an inspection of the network.  Here, we assume that if an alert happens on a link being used for exfiltration, the attack is successfully detected and stopped immediately.

While the defender chooses the alert threshold for each link, the attacker chooses the route and duration, $n$, of the exfiltration.  The exfiltration lasts for $n$ rounds, and each round $D/n$ data is exfiltrated.

Given the route, the distributions for each link, the thresholds, and the number of rounds, the probability of an exfiltration attempt remaining undetected in a single round is:
 \[
 P(\mathrm{undetected}) = \prod_{l \in route} \mathrm{cdf}\left(\mathrm{dist}_l, \mu_{l} + T_{l} \sigma_{l} - \frac{D}{n} \right).
 \]
This is the probability that a random variable drawn from each link's distribution, plus the extra data from the exfiltration, is less than the threshold---for all links in the route.

Given this probability, we can then calculate the expected reward for the attacker, $R_{att}$.  Exfiltration attempts are stopped immediately if detected, so the success of a round depends on the success of all previous rounds.  Attackers receive reward for any amount of data they exfiltrate, not just the complete data.  This gives us:
 \[
 E[R_{att}] = \frac{D}{n} \sum_{i=1}^{n} P(\mathrm{undetected})^i.
 \]

 On the defender's side, there are two types of cost: the cost of performing inspections after alerts, $C_{insp}$, and the costs associated with data loss, $C_{data}$.  Even in the absence of an attacker, links in the network will occasionally exceed the threshold, which will result in an ongoing base cost:
\[
 E[C_{base}] = C_{insp} \frac{\sum_{l \in links} |l_\mathrm{src}| |l_\mathrm{dest}| (1 - \mathrm{cdf}\left(\mathrm{dist}_l, \mu_l + T_l \sigma_l) \right)}{1-\delta}
\]
In Figure~\ref{fig:network1}, the Desktops and Internet-facing servers represent multiple devices (and therefore multiple network connections).  We treat these as identical in terms of network behaviour, which allows us to model large networks in an efficient way.  This has no impact on the attacker (any device would have the same probabilities, so additional devices can be ignored and do not need to be considered in routes), but it does have an impact on the defender's costs: each additional device can potentially raise alerts, increasing costs.  In the equation above, $|l_\mathrm{src}| $ and $|l_\mathrm{dest}|$ give the number of source and destination devices for a link, respectively.

The defender's base costs are ongoing---not just incurred while the attacker is acting---so we assume an infinite time horizon with a discount factor $\delta$.

To these ongoing base costs we can add the costs from data loss to arrive at the expected total loss for the defender:
\[
 E[C_{total}] = E[C_{base}] + C{data} \sum_{i=1}^{n} P(\mathrm{undetected})^i
\]

The defender's strategies consist of a choice of threshold for each link, and the attackers strategies consist of a choice of route and a choice of $n$, the exfiltration duration.  With the equations for attacker rewards and defender costs above we can now create the payoff matrices for any network configuration and set of strategies.  In the following sections we will first look at the characteristics of the model without strategic interaction, and then explore the strategic setting and Nash equilibria in different scenarios.

\section{Model Characteristics}

In this section we explore some of the characteristics of the model without considering strategic interaction.  First we look at how changing the detection threshold affects the system.  We look just at the exfiltration route Internal Server $\rightarrow$ Desktops $\rightarrow$ Remote Host.  The parameter settings are shown in Table~\ref{tab:parameters}.  In this illustration, we use the same detection threshold across all of the links, ranging from $1.5\sigma$ to $3.0\sigma$.

\begin{table}[h]
\begin{tabular}{@{}ll@{}}
\toprule
Parameter                                      & Value                       \\ \midrule
Data Loss Cost, $C_{data}$                     & 100000                      \\
Inspection Cost, $C_{insp}$                    & 100                         \\
Data Size, $D$                                 & 20.0                        \\
\multicolumn{2}{c}{Network Links}                                                  \\
Internal Server $\rightarrow$ Desktops         & Normal($\mu=5,\sigma=.5$)   \\
Internal Server $\rightarrow$ External Servers & Normal($\mu=15,\sigma=2.5$) \\
Desktops $\rightarrow$ Remote Host             & Normal($\mu=5,\sigma=.2$)   \\
External Servers $\rightarrow$ Remote Hosts    & Normal($\mu=10,\sigma=.25$)  \\ \bottomrule
\end{tabular}
\vspace{.5em}
\caption{Model parameters}
\label{tab:parameters}
\end{table}

\begin{figure}
    \centering
    \includegraphics[width=.92\columnwidth]{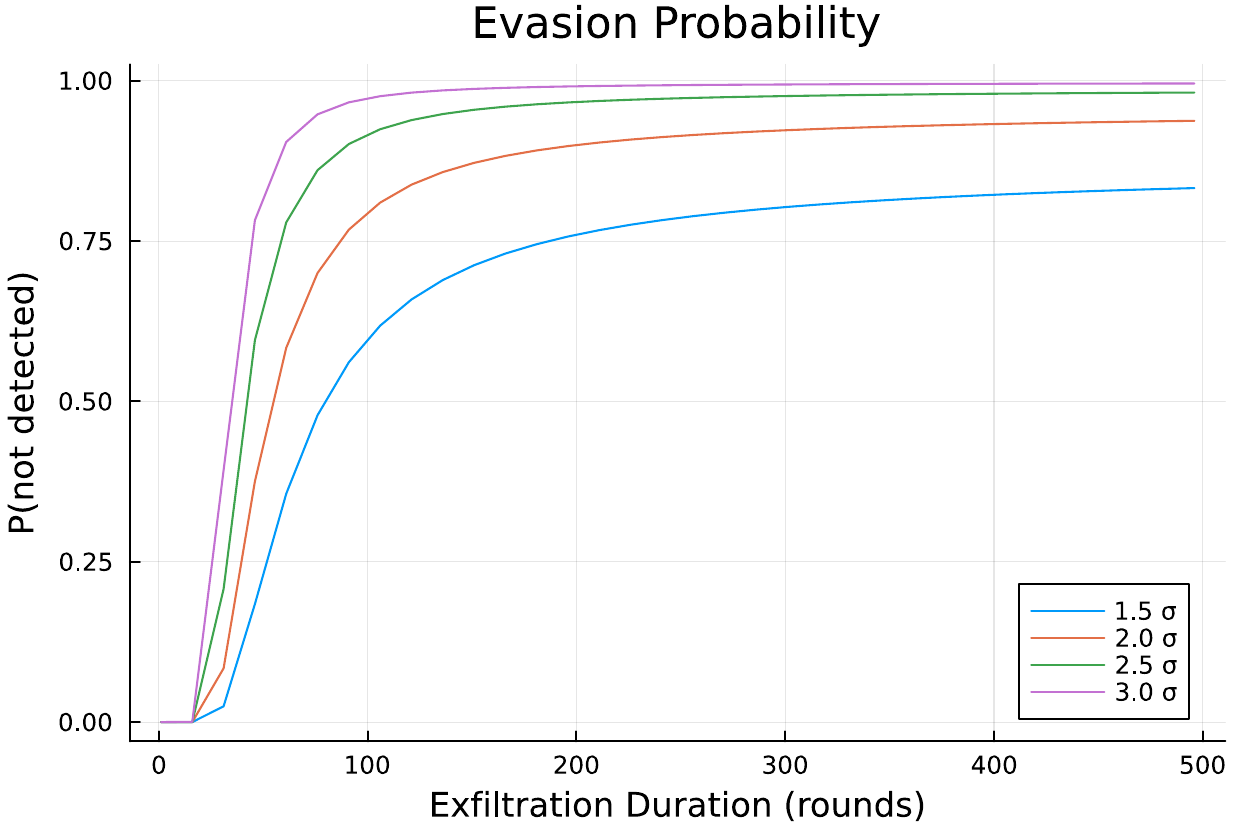}
    \caption{Probability that exfiltration is undetected (in a single round) for different detection thresholds.  Detection is more likely with lower thresholds.}
    \label{fig:evasion_probability}
\end{figure}

In the model, the attacker chooses the duration of the exfiltration.  The amount of data remains constant, so a longer exfiltration means that a smaller amount of data is being transferred each period.  As a consequence, shorter exfiltration durations, where more data is transferred each period, are more likely to be detected than longer durations.  This is visible in Figure~\ref{fig:evasion_probability}, which shows the probability \textit{per round} of the attack being detected for different durations and threshold values.  Higher detection thresholds lead to a higher probability of the exfiltration evading detection.

\begin{figure}
    \centering
    \includegraphics[width=.92\columnwidth]{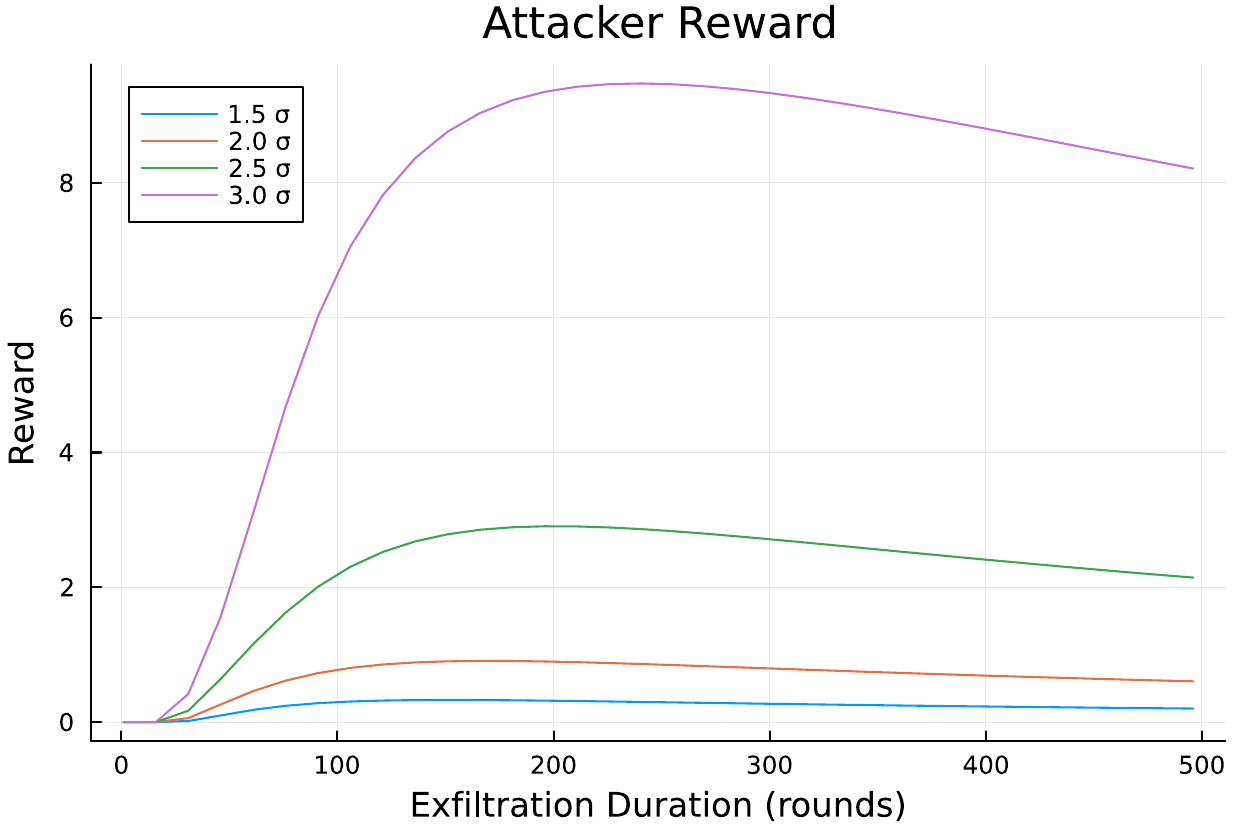}
    \caption{Expected payoff for the attacker given different detection thresholds.}
    \label{fig:attacker_reward}
\end{figure}

This naturally affects the payoff received by the attacker.  Figure~\ref{fig:attacker_reward} shows the expected reward for the attacker for different exfiltration durations and detection thresholds.  The reward is the amount of data the attacker successfully exfiltrates.  The maximum, in this scenario, is 20.  We can see in the figure that low thresholds successfully prevent almost all exfiltration; high thresholds allow much more data to be exfiltrated. 

\begin{figure}
    \centering
    \includegraphics[width=.92\columnwidth]{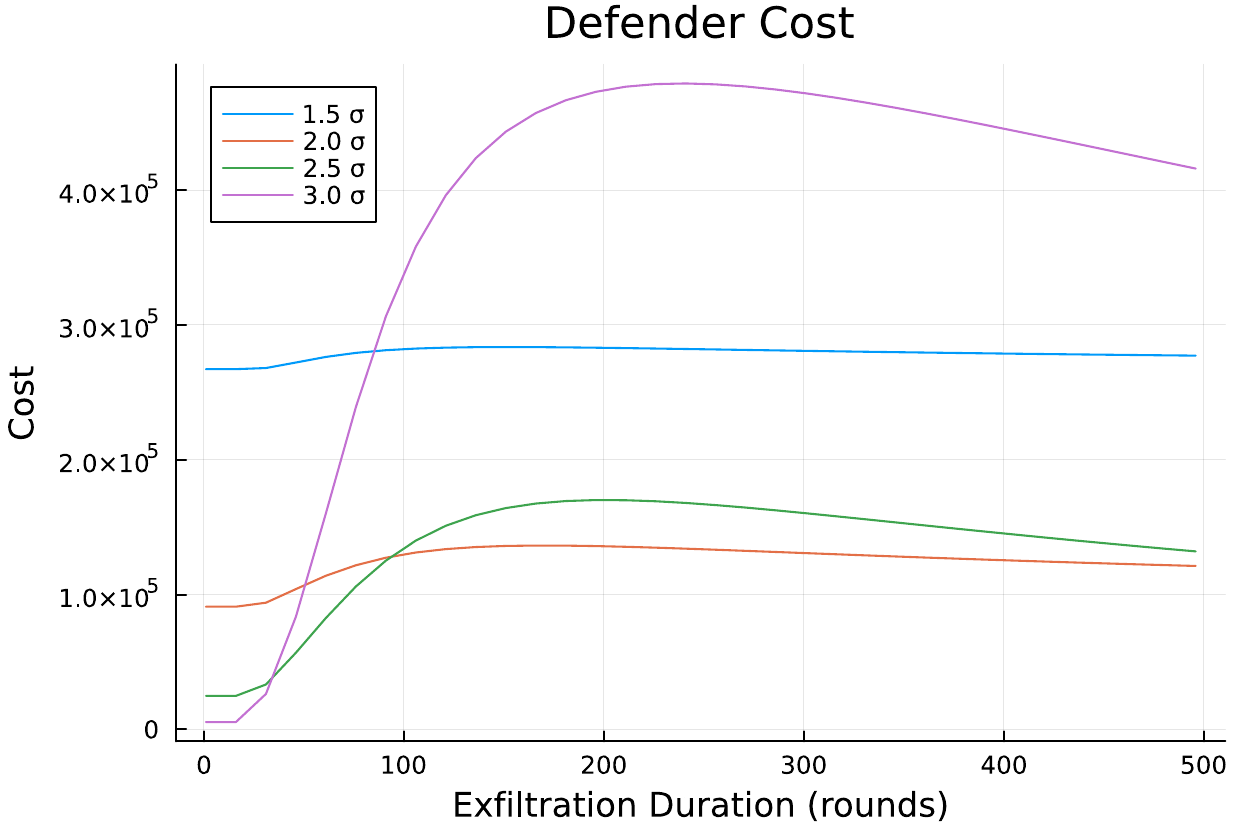}
    \caption{Expected total cost to the defender (inspection costs plus data loss costs) for different detection thresholds. }
    \label{fig:defender_cost}
\end{figure}

While low thresholds lead to very low amounts of exfiltration they also imply high levels of inspection due to false positives, which leads to additional cost.  The defender is trying to minimize total cost, which includes both costs from data loss and costs from network inspection.  The cost of too much inspection can be larger than the benefits gained from stopping exfiltration.  This can be seen in Figure~\ref{fig:defender_cost}: the lowest threshold, $1.5\sigma$, which prevents most exfiltration, has a very high constant cost.  On the other hand, high thresholds mean fewer inspection costs, but also lead to more costs from data loss, as can be seen with the $3.0\sigma$ threshold.  Values between these two extremes offer a balance between inspection costs and data loss prevention.  It is this value we seek to optimize in the presence of the attacker.

\section{Results}

We will now look at strategic interaction between the attacker and defender.  We start with the same parameters from Table:\ref{tab:parameters} in the previous section; however, we now allow the defender to set different thresholds for each link, and we allow the attacker to choose either route.

Allowing different thresholds for each link greatly expands the problem space, so we restrict the threshold options to $[1.8,2.0,2.2,2.4,2.6,2.8]$ and the exfiltration duration to $[1,10,20,35,40,\dots,350]$ to make a solution tractable. We solve for a mixed-strategy equilibrium and find that with these parameters, the defender has a cost of 919293 and the attacker has a reward of 1.1035.  The attacker's strategy selects route 2 approximately 67\% of the time, and route 1 the rest.  The defense strategy is also mixed.  Next we will explore how difference parameters affect the system.

\subsection{Data Size}

We start by looking at data size.  We find the Nash equilibria for models with various data size values while keeping other parameters the same.  We find that data size does not have any real impact on the system, except for the duration.  The attacker reward stays fairly constant, as does the share of route 2 in the mixed strategy.  The defense stays similar as well.


The duration does change---as data size goes up, duration increases from 40-50 to 280--350. Intuitively this makes sense: increasing the duration reduces the exfiltration rate, making detection harder.  The optimal behaviour for the attacker is to keep the rate relatively constant as the size changes, as it is the rate that triggers alerts.

\subsection{Link Variance}

Next we look at how the standard deviation of links can affect system performance.  In this model, the historical behavior of each network link is represented by a distribution; a lower variance or standard deviation means less variability from period to period, and a higher value means behavior is more variable.

Here, we set the values for both routes to be symmetric.  The left-side links (between the internal server and desktops/internet servers) will remain constant ($N(\mu=5,\sigma=0.5)$, and the standard deviation of the right-side links (connecting to the remote server) will be varied between $\sigma=0.25$ and $\sigma=2.0$, keeping both values the same, an with $\mu=10.0$.

\begin{figure}
    \centering
    \includegraphics[width=.92\columnwidth]{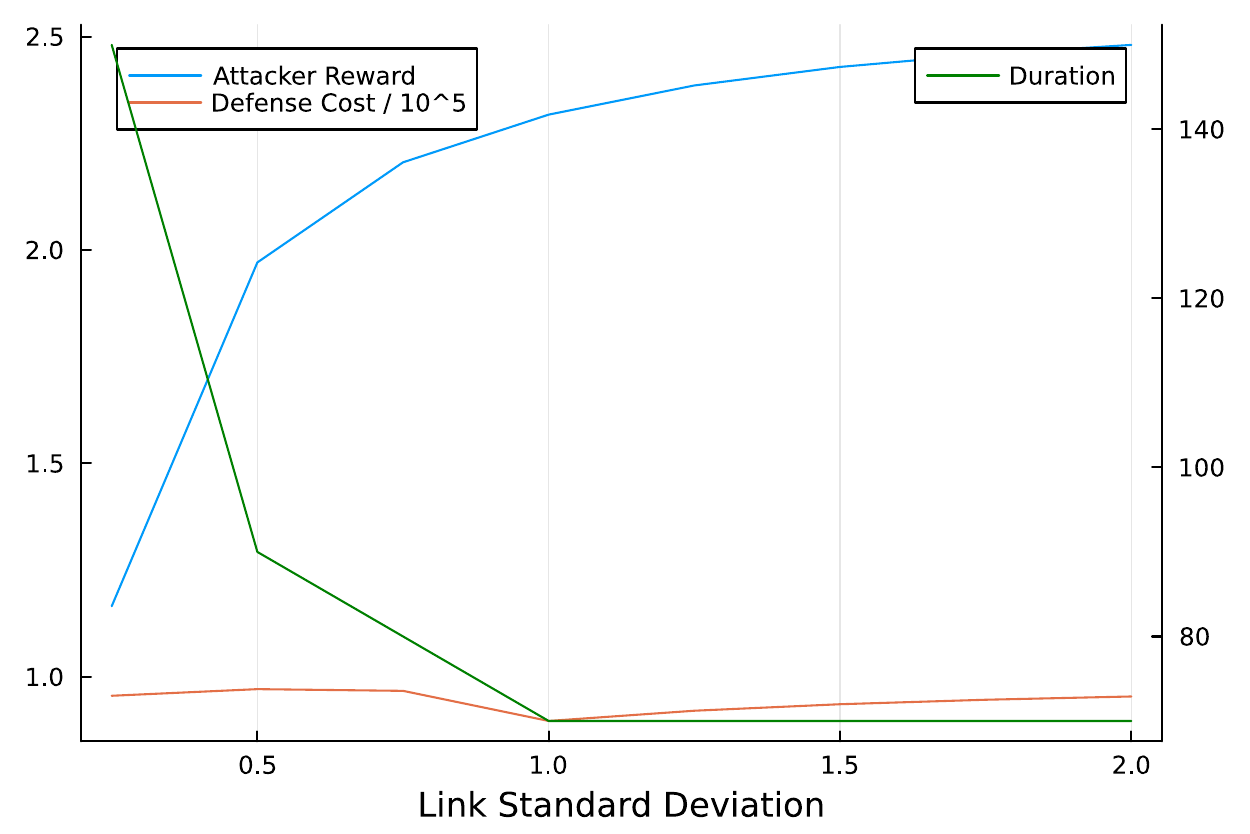}
    \caption{The effect of changing both the right-side links' standard deviations on system performance.}
    \label{fig:var_effect}
\end{figure}

The results are presented in Figure~\ref{fig:var_effect}, which shows three values.  The defense cost (scaled to fit on the axis) remains relatively constant, the attacker reward increases with the standard deviation, and the duration decreases as standard deviation increases.  As expected, the equilibria is symmetric: the defense strategy is split evenly between both routes, with thresholds of [2.2, 2.4, 1.8, 2.0] and [2.4, 2.2, 2.0, 1.8] for $\sigma=0.25$.  The order is [Internal$\rightarrow$Desktops, Internal$\rightarrow$External, Desktops$\rightarrow$Remote, External$\rightarrow$Remote]  For values $\sigma=0.5$ and greater, these thresholds are flipped to [1.8, 2.0, 2.4, 2.2] and [2.0, 1.8, 2.2, 2.4].  What this shows is that the optimal defense strategy is to put the lower thresholds where there is less variance---increasing the chance of detecting exfiltration, and reducing the chance of a false positive.

The attacker's strategy is also split evenly between both routes. With lower network variance, the attacker must use a longer exfiltration duration to lower the rate and avoid detection.  As the variance increases, faster exfiltration becomes optimal and a higher reward can also be achieved.

\begin{figure}
    \centering
    \includegraphics[width=.92\columnwidth]{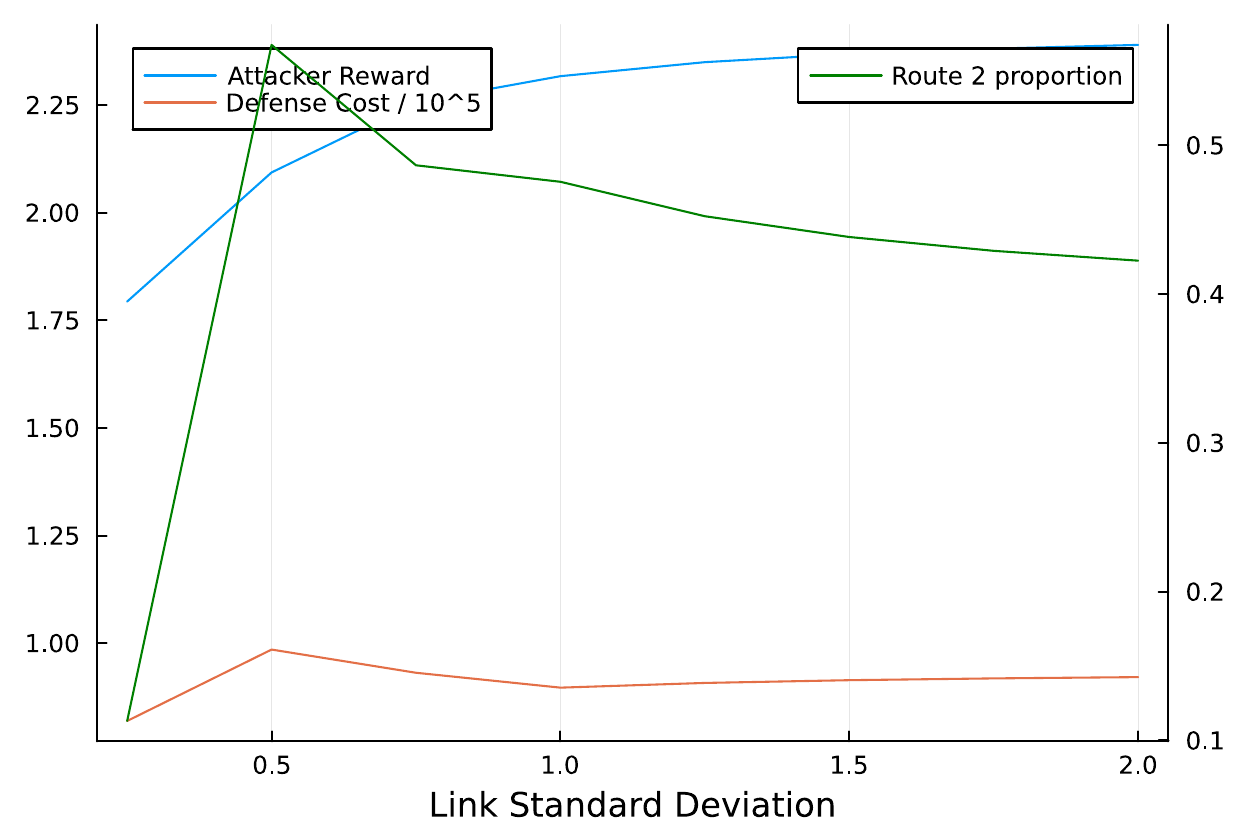}
    \caption{The effect of changing the standard deviations of just the External$\rightarrow$Remote link on system performance.}
    \label{fig:var_effect2}
\end{figure}

When the variance is split unevenly, the effect on the system changes (Figure~\ref{fig:var_effect2}).  The attacker reward still increases with variance, and the defense cost remains similar, but the weight of route 2 in the mixed strategy changes.  When the variance is very low, the strategy is weighted towards route 1---there is little opportunity for the attacker.  As variance is increased, the weight shifts in favor of route 2, but as the variance grows larger, this reverse back towards route 1.

\section{Conclusion}

This is work we have presented a simple data exfiltration game that can be used as a tool to explore the consequences of interactions between attackers and network defenders.  

While the scenarios discussed here were quite simple, this approach can be expanded to look at different network configurations as well as different types of exfiltration channels.  Additional links can be added to represent more network devices, and links with different properties can be used to represent different exfiltration methods.  For example, a link to a remote host with a very small mean could represent exfiltration over DNS, whereas links like those used in the examples above might represent HTTP exfiltration.

In this paper we used only Normal distributions, however others may be used and could potentially be a better fit for network behavior.  The use of distributions is convenient---network monitoring tools have more sophisticated ways to assess network behavior.  However, eventually these tools make a decision about whether or not the network exceeds some threshold and if some action should be taken, which we attempt to capture with our approach.  The standard deviation of a link distribution can also be interpreted as a measure of confidence in the ability to predict normal network behaviour.

One extension that might have an interesting effect on the results is adding a probability for correct detection when alerted.  That is, at the moment, alerts immediately stop exfiltration (if it is ongoing).  However, it is possible that after an alert an exfiltration attempt may go undiscovered, even after investigation.  In the current model, the attacker is probably incentivized to exfiltrate more quickly as any alert will halt the exfiltration.  In reality, many might be assumed to be false positives and ignored, giving the attacker additional time.

The problem space for these types of models can grow very quickly as extra devices and parameters are added.  Different approaches to solving for the equilibria could be employed, but structural approaches might be beneficial as well.  Thinking in terms of larger collections of devices and the interfaces between these groups might be one approach.  By looking just at the interface the complexity of the model and the problem space could be reduced make solutions more tractable.

\bibliographystyle{plain}
\bibliography{references}

\begin{thebibliography}{1}

\bibitem{durkota_optimal_2017}
Karel Durkota, Viliam Lisý, Christopher Kiekintveld, Karel Horák, Branislav Bošanský, and Tomáš Pevný.
\newblock Optimal {Strategies} for {Detecting} {Data} {Exfiltration} by {Internal} and {External} {Attackers}.
\newblock In Stefan Rass, Bo~An, Christopher Kiekintveld, Fei Fang, and Stefan Schauer, editors, {\em Decision and {Game} {Theory} for {Security}}, Lecture {Notes} in {Computer} {Science}, pages 171--192, Cham, 2017. Springer International Publishing.

\bibitem{mc_carthy_data_2016}
Sara~Marie Mc~Carthy, Arunesh Sinha, Milind Tambe, and Pratyusa Manadhata.
\newblock Data {Exfiltration} {Detection} and {Prevention}: {Virtually} {Distributed} {POMDPs} for {Practically} {Safer} {Networks}.
\newblock In Quanyan Zhu, Tansu Alpcan, Emmanouil Panaousis, Milind Tambe, and William Casey, editors, {\em Decision and {Game} {Theory} for {Security}}, Lecture {Notes} in {Computer} {Science}, pages 39--61, Cham, 2016. Springer International Publishing.

\bibitem{meurs_deception_2024}
Tom Meurs, Edward Cartwright, Anna Cartwright, Marianne Junger, and Abhishta Abhishta.
\newblock Deception in double extortion ransomware attacks: {An} analysis of profitability and credibility.
\newblock {\em Computers \& Security}, 138:103670, March 2024.

\bibitem{mitchell_game_2018}
Robert Mitchell and Brian Healy.
\newblock A game theoretic model of computer network exploitation campaigns.
\newblock In {\em 2018 {IEEE} 8th {Annual} {Computing} and {Communication} {Workshop} and {Conference} ({CCWC})}, pages 431--438, January 2018.

\bibitem{nguyen_stackelberg_2017}
Thanh Nguyen, Michael~P. Wellman, and Satinder Singh.
\newblock A {Stackelberg} {Game} {Model} for {Botnet} {Data} {Exfiltration}.
\newblock In Stefan Rass, Bo~An, Christopher Kiekintveld, Fei Fang, and Stefan Schauer, editors, {\em Decision and {Game} {Theory} for {Security}}, Lecture {Notes} in {Computer} {Science}, pages 151--170, Cham, 2017. Springer International Publishing.

\end{thebibliography}

\end{document}